\documentstyle[12pt, aasms4]{article}
\begin{document}

\title{The Early Palomar Program (1950-1955) for the Discovery of Classical Novae in M81:
 Analysis of the Spatial Distribution, Magnitude Distribution, and Distance Suggestion}

\author{Michael M. Shara} 

\affil{Space Telescope Science Institute} 
\affil{3700 San Martin Drive, Baltimore, MD, 21218, mshara@stsci.edu}
\affil{Astronomy Department, Columbia University}
\affil{Box 46, Pupin Hall, 538 West 120th Street, New York, NY, 10027}
\affil{Department of Astrophysics, American Museum of Natural History}
\affil{Central Park West at 79th, New York, NY, 10024}

\author{Allan Sandage}

\affil{Observatories of the Carnegie Institution of Washington}
\affil{Pasadena, CA, 91101}

\author{David R. Zurek}

\affil{Space Telescope Science Institute} 
\affil{3700 San Martin Drive, Baltimore, MD, 21218, mshara@stsci.edu}

\date{Recieved / Accepted}

\begin{abstract}

Data obtained in the 1950-1955 Palomar campaign for the discovery of classical novae in 
M81 are set out in detail. Positions and apparent B magnitudes are listed for the 23 novae 
that were found. There is modest evidence that the spatial distribution of the novae 
does not track the B brightness
distribution of either the total light or the light beyond an isophotal radius that is
$70\arcsec$ from the center of M81. The nova distribution is more extended than the 
aforementioned light, with a significant fraction of the sample appearing in the outer
disk/spiral arm region. We suggest that many (perhaps a majority) of the M81 novae
that are observed at any given epoch (compared with say $10^{10}$ years ago) are 
daughters of Population I interacting binaries. The conclusion that the present day novae
are drawn from two population groups, one from low mass white dwarf secondaries of 
close binaries identified with the bulge/thick disk population, and the other from
massive white dwarf secondaries identified with the outer thin disk/spiral arm 
population, is discussed. We conclude that the M81 data are consistent with the two
population division as argued previously from (1) the observational studies on other
grounds by Della Valle et al. (1992, 1994), Della Valle \& Livio (1998), and 
Shafter et al. (1996) of nearby galaxies, (2) the Hatano et al. (1997a,b) Monte 
Carlo simulations of novae in M31 and in the Galaxy, and (3) the Yungelson et al. (1997)
population synthesis modeling of nova binaries. Two different methods of using M81
novae as distance indicators give a nova distance modulus for M81 as $(m-M)_0 = 27.75$,
consistent with the Cepheid modulus that is the same value.

\end{abstract}


\section{Introduction}

A principal goal of the initial Palomar program on observational cosmology, that 
began with the commissioning of the 200-inch Hale telescope in 1949, was the 
testing and revision of the Mount Wilson extragalactic distance scale (Hubble 1951).
That scale was defined by Hubble's (1925, 1926, 1929) distances to NGC 6822, M33, M31,
and the galaxies immediately beyond the Local Group in the M81/NGC 2403 and M101 groups
(Hubble \& Humason 1931; Hubble 1936; Holmberg 1950).

An early central result was Baade's (1952) discovery that the RR Lyrae variables in the
disk of M31 did not resolve out of the background at the expected apparent magnitude of
$m_{pg} = 22.4$. Only the top of the globular cluster-like giant branch of the HR
diagram resolved at that level. By a series of arguments, Baade (1952, 1956) could show
that M31 was $\sim1.5$ mag further away than Hubble's modulus of $(m-M) = 22.0$, and that
the assumed zero point of the classical Cepheid period-luminosity relations was in 
error by about that amount.

A long-range program that was parallel to Baade's M31 compaign (Baade and Swope 1955, 1963)
was the study of the stellar content of other galaxies just beyond the Local Group, and 
also in selected E galaxies in the Virgo cluster. The purpose was to discover Cepheids
and novae in the M81/NGC 2403 and the M101 groups, and also to attempt discovery of novae
in the Virgo Cluster ellipticals.

Progress on this program was described in various yearly reports of the Mount Wilson
and Palomar Observatories (Bowen 1950-1970), and in the Introduction to the NGC 2403
Cepheid discovery paper (Tammann \& Sandage 1968). The final result of the
NGC 2403 campaign was that the distance modulus of that galaxy was $(m-M) = 27.56$
rather than Hubble's (1936) modulus of $(m-M) = 24.0$, giving a factor of $\sim5$
correction to Hubble's distance scale even at this very small distance beyond the
Local Group.

Other galaxies surveyed for Cepheids were M81 and M101, and less extensively for 
brightest stars in NGC 2366, NGC 2976, IC 2574, NGC 4236, Ho I, Ho II of the M81/NGC 2403
Group (Sandage and Tammann 1974a), and NGC 5204, NGC 5474, NGC 5477, and NGC 5585 and 
M101 itself in the M101 Group (Sandage and Tammann 1974b). A progress report was given
by Sandage (1954).

After NGC 2403, the most complete coverage for Cepheids and normal novae was in M81, 
considered by Hubble, and assumed on that basis by Holmberg (1950), to be at the same
distance as NGC 2403.

The Cepheid program for NGC 2403, M81, and M101 was moderately telescope-intensive
from 1950 through 1955. To assure somewhat adequate coverage for both Cepheids and novae,
the observing runs during the two weeks of dark of the moon were usually split into three
intervals of two days at the beginning of the 14 day interval, two days near the middle
and two days at the end. By the end of 1955, a total of 79 blue plates had been taken of
M81. The principal observers were Humason (30 plates), Sandage (38 plates), Baade 
(5 plates), Baum (3 plates), Hubble (2 plates), and Minkowski (1 plate).

Until his death in 1953, Hubble blinked the total material available at the time. He
discovered 10 faint variables (all near the plate limit at B$\sim23$), and 18 
classical novae in M81. The program was continued after Hubble's death so that at the 
end of the campaign in late 1955, a total of 23 novae, 30 suspected faint variables
(many of which are undoubtedly Cepheids), and 7 luminous blue variables (LBVs) had been
discovered in M81. Two novae were also found in the E galaxy NGC 4486 in the Virgo cluster
(Bowen 1952; Pritchet \& van den Bergh 1987).

None of the detailed data on the M81 novae or faint variables has been published.
However, in the 1954 summary mentioned above, Hubble's preliminary result on the modulus
of M81, based on the first 18 novae, was discussed. Using a provisional apparent
magnitude scale that had been set up by one of us (AS) before the photoelectric
magnitude sequences had been established in the 1960's in NGC 2403, M81, and the M81
companion of Ho IX, Hubble had concluded by early 1953 that M81 is $\sim3.8$ mag
further away than M31.

In a remarkable procedure, Hubble reduced the novae data that were available to mid-1953
to the mean apparent magnitude of the nova system, averaged at 14 days after maximum.
He used the number of days that had elapsed between the discovery date of a given nova
and the last previous plate of the galaxy. In this way he calibrated the ``dead time
correction'' to an inferred maximum magnitude by the statistical properties of the nova
system using his 86 novae in M31 (Hubble 1929) as a template. The procedure was 
approximate at best (although similar to the ``control time'' algorithm now used
extensively). Furthermore, no account was taken of the absolute maximum magnitude-decay
rate (MMRD) relation of novae, found earlier by McLaughlin (1939, 1945, 1946) and fully
confirmed and extended by Arp (1956), Schmidt (1957), Rosino (1964), Shara (1981a,b), 
Cohen (1985), Capaccioli et al. (1989), among others. Remarkably, however, the absolute
magnitude of both ``fast'' and ``slow'' novae are now known to be closely the same at 
14 days after maximum. The different shapes of the light curves all cross in a composite
light curve near this time from maximum (eg. Buscombe \& de Vaucouleurs 1955; Shara 
1981a).

Data on two of the Cepheids (V2 and V30 in an internal numbering used in the original
working identification charts) were also analyzed (Sandage unpublished). The result
was that periods were determined to be 30.073 and 30.625 days with mean B magnitudes
of 22.5 and 22.6 respectively. Using these variables, Freedman and Madore (1988) 
measured I magnitudes for them. They derived an M81 modulus of $(m-M)_0 = 27.59$
(see also Freedman et al. 1994). This modulus, combined with the original Palomar
modulus of NGC 2403 (Tammann and Sandage 1968), confirmed the assumption of Hubble
and Holmberg that M81 and NGC 2403 form a group at closely the same distance. The
Freedman/Madore data also corrected a late, aberrant, claim to the contrary (Sandage 1984)
that $(m-M) = 28.8$ for M81 that was based on a false precept concerning the M81 data,
as one of us (AS) unfortunately set out in 1984.

The purpose of the present paper is to publish the data for the Palomar M81 novae. These
data also permit discussion of the implications of the surface distributions of the novae
over the face of M81, compared with similar data for M31, for a division of normal 
novae into at least two classes, both spectroscopically (Williams 1992, Della Valle \&
Livio 1998), and spatially in the Galaxy and in M31, M33, and LMC (Della Valle et al.
1992, 1994). This division into separate populations is now widely believed to be caused
by a difference in the mass distribution of the white dwarf progenitors to the novae,
as discussed in section 4.

\section{Novae as Distance and Binary Star Population Indicators}

With the discovery of the eclipsing light curve of the old nova DQ Her
(Walker 1954, 1956) and the susequent discovery of periodic radial velocity variations
in the many old and recurrent novae (Kraft 1964), and based on the mass transfer model
for the U Gem cataclysmic variable AE Aqr (Crawford \& Kraft 1956), Kraft (1959, 1963,
1964) argued that all normal novae occur in close binary systems. Walker's (1968) 
subsequent discovery that the classical nova T Aur is also an eclipsing binary added
to the evidence. ``The model, in which a late type star is losing mass through the 
inner Lagrangian point to a compact companion, has become standard for 
cataclysmic variables'' (Robinson 1976, Warner 1976). It has also become standard for
normal novae (Gallagher and Starrfield 1978; Shara 1989 for reviews).

The physical process leading to the large energy release in the outburst is known (with
almost definitive certainty) to be a thermonuclear runaway caused by the ignition of 
hydrogen (burning into helium) after a critical mass is reached of accreted gas from the 
primary onto the surface of the white dwarf via the accretion disk.

The model was developed over a three decade period by a number of authors. Entrance
to the extensive early literature can be made through the defining papers of the process
by Schatzman (1949, 1965), Starrfield, Sparks, \& Truran (1975, 1976), Sparks, 
Starrfield, \& Truran (1977a,b), Prialnik, Shara, \& Shaviv (1978, 1979) with references to the 
many other principal authors
therein. More recent papers and reviews are by Shara (1989), Truran (1990), Livio (1992),
Della Valle (1992), Della Valle \& Livio (1995), and papers to be cited later herein, as
well as the recent monograph of Warner (1995).

Because all normal novae are close binaries with mass exchange, it is clear that novae 
will erupt in all galaxies at all epochs after which close, mass-exchange binaries,
one of which is a white dwarf, have been formed. The thermonuclear runaway occurs when
the degenerate hydrogen that has been accreted onto the white dwarf surface from the 
Roche-lobe secondary is compressed beyond critical density on the surface. The
thermonuclear heating from the nuclear reactions relieves the degeneracy, leading to
rapid expansion and expulsion of the white dwarf envelope. An Eddingtion (or even
super-Eddington) photon flux ensues, with an eventual decline of the light curve as
envelope exhaustion occurs. Because the physics is well understood (Shara 1981a,b, 1989;
Livio 1992), and because the novae luminosities are so high at maximum light, simply
the discovery of novae in external galaxies provides a powerful method to trace the
close binary and white dwarf star populations in the parent galaxies. The division into
two nova groups, depending on the mass of the white dwarf component, also provides a
method to study the different evolutionary properties of the older bulge/thick disk
and the younger outer thin disk/spiral arm populations where the mass spectrum of the 
white dwarfs is expected to be different.

\section{The M81 Novae Data}

\subsection{The Observing Record}

We list in Table 1 all 5 meter Palomar plates taken for this program, including those on
which no M81 novae appear. The table contains plate number, observer, date taken, Julian
date, plate quality and the novae visible on each plate.

\subsection{Photometry}

Magnitudes of the novae on the plates of M81 where they appear were determined using 
local magnitude sequences (not shown) that were set up near each nova or groups of
adjacent novae. The sequences were transferred, and combined, from three separate master
photoelectric sequences that had been determined earlier in other programs. These 
master sequences were in Selected Area 57, (unpublished but used extensively since 1952,
based on data from a number of Mount Wilson and Palomar observers; see eg. Majewski, 1992,
and Reid \& Majewski, 1993). The two other primary photoelectric sequences are in 
NGC 2403 (Tammann \& Sandage 1968), and Ho IX and M81 itself (Sandage 1984).

The systematic reliability of the latter two sequences, at the level of 0.1 mag, have 
been verified by Metcalfe \& Shanks (1991). Independent verification of the Ho IX
master sequence, also to this level, was made by one of us (MS) with CCD images kindly
supplied by George Jacoby. The accuracy of the transferred secondary sequences that were
spread over the face of M81 was also tested by Judith Cohen (1984 unpublished) with the
result that our secondary sequences here, upon which our nova photometry rests, have been
confirmed in systematic accuracy to a level of 0.2 mag. This is sufficient for the
present purposes.

The B magnitudes of each nova visible on each plate, measured relative to the local
magnitude sequences just described, are listed in Table 2.

\subsection{Astrometry}

The positions of the 23 novae have been measured from the discovery plates using the
two axis Grant machine at the Kitt Peak National Observatory. The B1950 and J2000
coordinates and the apparent radial and de-projected radial distance of each nova 
is listed in Table 3.

Calculating the de-projected radial distance requires that we know the inclination and
the orientation of M81 on the sky. Isophotes were fit to M81 using the program ellipse 
in the stsdas.analysis.isophote package with in IRAF. This process also determines the 
ellipticity and the position angle (measured clockwise from the Y-axis of the CCD image)
of each isophote. The inclination can be determined using the relationship 

$$cos^2(i) = \frac{(\frac{b}{a})^2 - r_{o}^2}{1-r_{o}^2} $$

\noindent
(Tully \& Fisher 1977; see also Hubble 1926 and Sandage et al. 1970) where $r_o = 0.2$ (this is the 
assumed axial ratio for a system completely edge on). Since, $e = 1 - \frac{b}{a}$ and the ellipticity 
from the isophote fitting is 0.4662 we find that $\frac{b}{a} = 0.5338$. It follows that $i = 60.4$ using
the above equation. Tully and Fisher (1977) found $i = 58$ which is in excellent 
agreement with our value. 

To fully correct for the inclination of the galaxy the positions of the novae must be
placed in the coordinate system of the galaxy and then corrected for the inclination. We
have chosen the major axis as our X-axis and the minor axis as our Y-axis. The position
of each nova is determined in the following way

$$ X = N_{ra} cos(\theta) - N_{dec} sin(\theta) $$
$$ Y = \frac{N_{ra} sin(\theta) + N_{dec} cos(\theta)}{cos(i)} $$

\noindent
$\bullet~~  N_{ra}$ is the distance of the nova from the center of M81 in arcseconds of right 
ascension ($N_{ra} = 15 cos(M81(DEC)) (Nova(RA) - M81(RA))$).

\noindent
$\bullet~~  N_{dec}$ is the distance of the nova from the center of M81 in arcseconds of 
declination ($N_{dec} = Nova(DEC) - M81(DEC)$).

\noindent
$\bullet~~  \theta$ is the angle of the major axis from West (-121.44 degrees).

\noindent
The distances in X and Y are added in quadrature to give us the corrected radial distances
for each nova presented in Table 3.

\section{Spatial Distribution of Novae in Galaxies}

\subsection{Two Populations of Novae}

A strong debate on the spatial distribution of novae in galaxies and the populations to
which they belong has appeared in the literature during the past decade. Ciardullo et al.
(1987) and Capaccioli (1989) supported the view that most of the M31 novae are produced
in the galaxy's bulge. However, Della Valle et al. (1992, 1994, their Fig. 3), in
discussing the distribution of Galactic novae relative to their distances above the 
galactic plane, and also concerning the frequency of novae in late type galaxies
(M33, LMC) compared with bulge-dominated galaxies, concluded that younger, blue
populations (outer disk and spiral arm regions and the fraction of bulge population
that is young) produce {\it most} of the novae per unit K-band luminosity in all 
galaxies, regardless of Hubble Type. (We note, however, the recent criticism of the 
Della Valle et al. (1994) nova rates (because of normalization problems) by Shafter, 
Ciardullo, \& Pritchet (1999).) Della Valle et al. (1992,1994) showed that the division into
the two population groups is also supported by the difference in the decline rate
distributions of the light curve between early and late type star-producing galaxies,
ie. M31 vs LMC and M33, (Della Valle et al. 1994, their Figs. 1 and 3). 

The supposition is that the brighter, faster novae are in the young population where the
white dwarf progenitor is expected to be of higher mass than in the older population.
This is because the main sequence star that becomes a white dwarf in the outer disk
and spiral arm populations presumably is of higher mass when it leaves the main
sequence than progenitor stars in the bulge/thick disk population, at least at the 
present epoch.

This follows because there is a strong relation between the final white dwarf mass and
the initial mass of the original star. The higher the initial mass, the higher will be
the white dwarf remnant after evolution. This is the famous initial mass-final mass
relation for white dwarfs, now apparently solved beyond credible doubt (Weidemann \&
Koester 1983, Fig. 1; 1984; Weidemann 1990), based in part on the central discovery
of massive white dwarfs in the young Galactic cluster NGC 2516 (Reimers \& Koester
1982), together with later discoveries of the same type.

Because the luminosity of the nova outburst is a strong function of the WD mass (going
as the cube of the mass; Shara(1981b), Livio 1992), the strength of the outburst and the
decay rate of the light curve are expected to differ according to the mass of the
envelope-exploding white dwarf. The spectroscopic differences between fast and slow novae
(Williams 1992; Della Valle \& Livio 1998), and the striking difference in the ejection
velocities summarized in these papers (eg. Fig. 2 of the last reference) are also 
explained in this way.

Observations of the novae in M51, M87, and M101 support this view (Shafter, Ciardullo, \&
Pritchet 1996). Furthermore, the nova {\it rate} per unit mass in a young, blue, stellar
population is expected to be higher than in an old, red population. Yungelson, Livio, \&
Tutukov (1997) predicted that this is because the massive white dwarfs produced in the
young population need only accrete hydrogen from their companions for a relatively
short time to reach the critical envelope mass and erupt as novae. These authors also 
suggest that the apparent numerical dominance of Galactic bulge novae over Galactic
disk novae is an observational selection effect: disk novae are more likely to be dimmed
by dust than bulge novae, therefore, apparently reducing their observed frequencies.
Monte Carlo simulations by Hatano et al. (1997a,b) on novae in M31 and in the Galaxy 
strongly suggest the above selection effect, and show the possibility for a
true dominance of disk novae over bulge/thick disk novae. (The Hatano et al. result depends
on the accuracy of their light plus dust model for M31, which still requires verification.)
Nova-rate studies in galaxies of different Hubble type (Della Valle et al. 1994, their 
Fig. 3) support this view.

\subsection{The M81 novae}
 
The apparent distribution of the 23 novae over the face of M81 is shown in Figure 1, 
overlaid on a KPNO service CCD image of the galaxy. The absence of the novae within
$70\arcsec$ of the center (the ``nova hole'') is similar to that found by Hubble (1929), Arp
(1956), and in the Asiago survey (Rosino 1964; Capaccioli et al. 1989). It is almost
certainly due to discovery-incompleteness in the broad-band B surveys (eg. Ciardullo
et al. 1987) near the center. The faintest detections of the M81 novae are at $B \sim 23.0$.
We also note the detection at $B=22.9$ of nova \#15, one of the novae closest to the center
of M81. 

The seven of the 23 novae in M81 that we have found in the central part of the M81 bulge
are numbers 7,9,11,14,15,18, and 24. These could be associated with the low mass white
dwarf bulge population. However, as suggested by the Hatano et al. (1997a,b) 
simulations, many of the {\it apparent} bulge novae must also belong to the young
spiral population. We note without comment that dusty spiral arms in M81 do in fact 
extend all the way into the central region of the M81 bulge (Fig. 2).

The results of the Hatano et al. simulations are as follows. Assuming a range of 
bulge-to-disk novae and adopting the observed distribution of Galactic classical
novae, Hatano et al. found that at least 67\% (and more likely 89\%) of the Galactic
novae belong to the disk. They found a similar result for the M31 novae.

Is the observed distribution of M81 novae in Figure 1 consistent with this finding? 
Consider first the standard method of analysis, used in many of the cited prior 
studies via the method of cumulative spatial and light distributions.

Figure 3 shows the data in Table 2 analyzed in several ways. The M81 B band isophotal
light is shown as the dashed curve. The isophotal light outside the isophote at $70\arcsec$,
the radius where our plates begin to detect novae, is shown as the dotted (lower) 
light curve. The largest difference between the ($70\arcsec$) isophotal light and the 
{\it deprojected} nova radial distribution is $D=0.23$. The Kolmogorov-Smirnov statistic 
then states that the novae do not follow the galaxy light, but only with 80\% confidence. 
Deprojection of the novae is only meaningful, of course, if the novae belong to the disk
population. If the novae belong largely to the bulge population than figure 3 supports
the view of Moses \& Shafter (1993) that the distributions of light and novae in M81 are
the same.

The second demonstration that our sample contains many disk novae is the correction for
incompleteness. As noted above, we have not detected the novae within $70\arcsec$ of the 
center. Using the Ciardullo et al. (1987) data with their much more complete survey of
the center of M31 to detect novae using narrow band H alpha emission rather than
broad-band continuum light, we can use the comparison in the Ciardullo et al. data of
the number of their detected novae in and outside the central region to calculate our
incompleteness.

The portion of M31 surveyed by Ciardullo et al. covered an area of 15 by 30 arc minutes
along the minor and major axes of that galaxy respectively. M81, with $(m-M)_0 = 27.75$ 
is $\sim4.5$ times farther away than M31 with $(m-M)_0 = 24.4$. Hence, the region
surveyed in M81 that would be equivalent to that in M31 is 3.3 by 6.7 arc minutes 
along the M81 minor and major axes.

In their complete H alpha survey, Ciardullo et al. found a total of 35 novae, of which
21 were within 5 arc minutes of the center of M31. This distance corresponds to the 
$70\arcsec$ radius of the ``nova hole'' in our photographic survey of M81 where we found no 
novae. Because the Ciardullo survey is beyond doubt virtually complete, whereas
our broad-band survey, as in Hubble, Arp, and the Asiago (Rosino), is not, the
Ciardullo ratios should closely define our incompleteness factors. Thus, we expect
that we have missed $\sim 21/35 = 60\%$ of the novae in the central 3.3X6.7 arc minute
region of M81. We did find 11 objects in this area. Therefore we must have missed
$\sim17$ objects during the survey time.

We also find that 13 of our 23 detected M81 novae lie {\it outside} the central
3.3X6.7 arc minute region. If we assume that we are complete in the discovery in
this region, our complete survey, adding the $\sim17$ novae assumed missed in the
``nova hole'', should have detected $(17+11+13) = 41$ novae. Having missed 17, we 
conclude that our {\it total} M81 survey is $17/41 = 41\%$ incomplete. Expressed the
other way, it was $23/41 = 59\%$ complete.

The incompleteness factor has been accounted for in Figure 4, which is the same as
Figure 3 but with the 17 assumed novae missed in the central $70\arcsec$ radius
added, assuming a radially uniform distribution of the undetected 17 central novae.
Neither this radial distribution, nor any other addition of 17 inner novae can
bring the galaxy isophotal light and the nova radial distribution into agreement in
Fig. 4.

There is, however, yet another central fact in the argument. Figure 5 is the same as
Figure 1 but with the area surveyed by Ciardullo et al. marked, showing that they could
not have found novae in the outer disk and spiral arms in M31, novae that are 
unquestionably of the arm (high mass white dwarf progenitor) population. Using our
statistics of 13 spiral-arm-population novae in M81 out of a total (completeness
corrected) of 41 novae of both population types, we would expect that Ciardullo et al.
have missed $13/41 = 31\%$ of the total M31 nova population, almost all of which will
be of the disk/spiral arm type.

From the above arguments concerning completeness, we conclude that {\it at most} 
$(41-17)/41 = 59\%$ of all the novae in M81 are in the bulge. Given (1) the small
number statistics, (2) the uncertainties in the dust models of Hatano et al. (1997a,b),
(3) possible differences between the M81 and M31 dust and nova distributions, and 
(4) the assumption that the limiting absolute magnitude survey limits for M31 and M81 
are similar, we cannot claim a stronger value for the bulge/arm ratio. However, Figures 3 and 4
and the work of Hatano et al. support the value we have given that supports an 
appreciable outer disk/spiral arm nova population that is consistent with the 
two-population dichotomy of Duerbeck (1990), Della Valle et al. (1992, 1994), Williams
(1992), and others cited in the above discussion.

It must be mentioned that the surveys of M31 by Arp and by the Asiago group 
(eg. Rosino 1964) cover a much larger region than the survey of Ciardullo et al., 
reaching to greater than 30 arc minutes radius from the center, therefore
encompassing more of the M31 spiral pattern. There is indeed evidence, first set
out by Arp (1956, his Fig. 36), for a bimodal distribution of magnitude at maximum
(see also Della Valle \& Livio 1998, their Fig. 4, which also is from Arp) for the 
M31 novae. This distribution is also separated into the fast and slow groups, as
are the characteristics of the novae in early and late type spirals (Della Valle et al.
1994, their Fig. 1).

\section{A Nova Distance to M81}

The Palomar survey of M81 for novae was not sufficiently dense to unambiguously 
determine the novae magnitudes at maximum light, nor the decay rate for the MMRD 
relation (eg. Arp 1956; van den Bergh 1975; Della Valle et al. 1994) that is 
necessary to determine a nova distance, with one notable exception. Table 2 for the
photometry shows that nova \#15 was discovered within one day of maximum and was 
followed for 35 days thereafter, giving a good determiniation of the decay rate.
Figure 6 shows the light curve. The least-squares slope using the five observed
data points is $0.0461 \pm 0.0061$ B mag/day. This corresponds to a time of decline by
three magnitudes of $t_3 = 65.1$ days. The MMRD relation, calibrated elsewhere
(Shara 1981b), is $M_B(max) = -10.1 + 1.57 log t_3$, giving $M_B = -7.25 \pm 0.10$
for nova \#15. From $B(max) = 20.6$, and using an estimated absorptions of 0.1 mag
(Sandage \& Tammann 1987, column 13, noting that nova \#15 is in the bulge and 
is assumed to suffer no internal absorption within M81), or 0.41 mag (Peimbert \& 
Torres-Peimbert 1981) gives $(m-M)_0 = 27.75$ or $(m-M)_0 = 27.44$. The former is
identical to the M81 Cepheid distance (Freedman et al. 1994). The true uncertainty 
using nova \#15 is, of course, at least as large as 0.31 mag because of the highly 
uncertain absorption.

An independent estimate of the nova distance to M81 is by comparison of the {\it
distribution} of the magnitudes in Table 2 with similar data for novae in M31. The
brightness distribution in the two galaxies to the completeness limit of the novae 
in M81 are expected to be similar if, (1) the sampling frequencies are similar,
(2) the sample sizes are similar, and (3) the novae are drawn from the same 
populations. (Note that a potential fourth factor - metallicity - is ignorable. 
Nova eruptions are independent of the metallicity of the parent galaxies and/or 
the nova environment because the enrichments of CNO and the O-Ne-Mg
elements by factors of between 5 and 50 are produced by hydrogen envelope
mixing with the underlying white dwarf (Shara 1981b)).

Each of the three requirements appear to be approximately fulfilled in the case of 
the M81 program compared with the M31 program of Ciardullo et al. (1987). (1) For both
galaxies, a few observing runs of several nights length typically occurred each
year, so the sampling frequencies are similar. (2) The 35 novae from Ciardullo et al.
are only $\sim1.5$ times more numerous than the 23 novae in the present survey - a 
difference than is not very significant in this context. (3) As we noted in section 4,
our M81 nova sample is incomplete in the central part of the galaxy. The M31 survey
of Ciardullo et al. is incomplete in the outer part of the galaxy, dominated by disk
novae. However, if disk novae are equally significant in both galaxies {\it even
in the bulge}, then the brightness distributions of the total samples are likely to
be similar. Nevertheless, large, complete magnitude-limited samples of novae over
the entire extents of M81 and M31 are required to confirm the results below.

These apparent (ie. as actually observed in the discovery programs) brightness 
distributions for both galaxies are plotted in Fig. 7. The top distribution is from
the M31 Ciardullo et al. sample. The bottom is from our M81 data in Table 2.

Two important features of these distributions are useful in comparing the two 
samples. (1) The brightness of the single brightest nova in each sample is indicative
of the most massive (disk) white dwarf in each sample. (2) The rapid increase
in the number of nova detections that are, say, 1.4 mag fainter than the brightest
nova is likely to be due to the ease of discovering these novae which must be 
near-Eddington luminosity objects that remain close to their maximum brightness 
for several weeks.

Figure 7 shows that, on the precepts set out above, the differential distance modulus
between M81 and M31 is $\sim3.4\pm0.3$ mag (remember Hubble's value of 3.8). Adopting
the true (absorption free) modulus for M31 as $(m-M)_0 = 24.26$ from IR photometry
(Welch et al. 1986), and assuming similar modest internal absorptions (0.3 mag) in the 
bulge of each galaxy, we derive $(m-M)_0 = 27.6 \pm 0.3$ for M81. Although far from
definitive, this value is in good agreement with the MMRD distance derived above 
from M81 nova \#15 as $(m-M)_0 = 27.75$, and the Cepheid distance (Freedman et al. 1994)
of $(m-M)_0 = 27.8$.

As to the general use of novae in external galaxies as distance indicators, we only
note here, as have others, that the brightest novae are $\sim3$ mag brighter than
the ``average'' Cepheids. Furthermore, the physics of the nova eruption is
now understood via the well defined MMRD relation, both from theory (Shara 1981a,b, 1989;
Livio 1992), and from observation (McLaughlin 1945; Arp 1956; Cohen 1985; Della Valle
et al. 1994, and many others). Therefore, because there are now precedures to make use
of novae (here, and eg. Pritchet \& van den Bergh, 1987), there is no question that
surveys of normal novae in distant galaxies, done with understanding of the novae
systematics now known, will be one of the more important observational programs in the
future that will help to carry the quest for the local extragalactic distance
scale to completion.

\acknowledgements
{We thank George Jacoby and Debra Wallace for obtaining KPNO CCD images
of M81 to permit us to test the M81 photoelectric sequences that 
had been set up by AS previously in Ho IX and M81 itself at Palomar. 
AS thanks Judith Cohen for her independent (unpublished) testing using
CCD technology of many of the local magnitude sequences over the face of
M81 in 1984. MS thanks Ed Carder for setup assistance at the KPNO two axis 
Grant measuring engine, and Mike Potter of STScI for assistance with data reductions.
John Bedke made the reproductions of the original Hubble/Sandage finder charts,
for which we are grateful. The Mount Wilson/Palomar commitment of the early
``nebular group'' of that Observatory to the M81 nova campaign in the first
decade of the 1950s at Palomar is evident. In that regard, AS is grateful
to the Palomar mountain crew, from night assistants to all mountain personnel,
for their crucial work behind the scenes in the observing period in that heady
epoch nearly 50 years ago in which the data that are discussed here were obtained.}

\newpage
\centerline{Figure Captions}
\parskip 2mm
\parindent 0mm

\figcaption{
The positions of the 23 novae discovered in the five year Palomar photographic
survey (1950-1955) with the 200-inch Palomar Hale reflector. Position data are
in Table 3.
}

\figcaption{
The central $\sim2'$ of M81 taken with the F547M filter and WFPC2 on HST. 
We have shifted the image by 5 pixels in each of X and Y and differenced
the image from itself to highlight faint features.
Note that the spiral dust lanes and the population indicators are detectable
almost to the nucleus of the galaxy.
}

\figcaption{
The cumulative radial number distribution of the 23 discovered novae and the 
isophotal light in M81. 
}

\figcaption{
Same as Fig. 3 but with the assumed 17 that we missed in the survey added
in the inner $\sim70''$ radius central regions. see text for details
}

\figcaption{
Same as Fig. 1 but with the 3.3X6.7 arc minute survey field of Ciardullo et al. 
(1987) superposed.
}

\figcaption{
The light curve of Nova \# 15 in M81 (data in Table 2). The dashed line defines
a decay rate of $0.0489 \pm 0.0076$ B mags/day.
}

\figcaption{
Brightness distributions (at the discovery apparent magnitudes) of the actual
discoveries of novae in M31 (top) from the survey of Ciardullo et al. (1987), 
and (bottom) from our M81 survey here (Table 2).
}


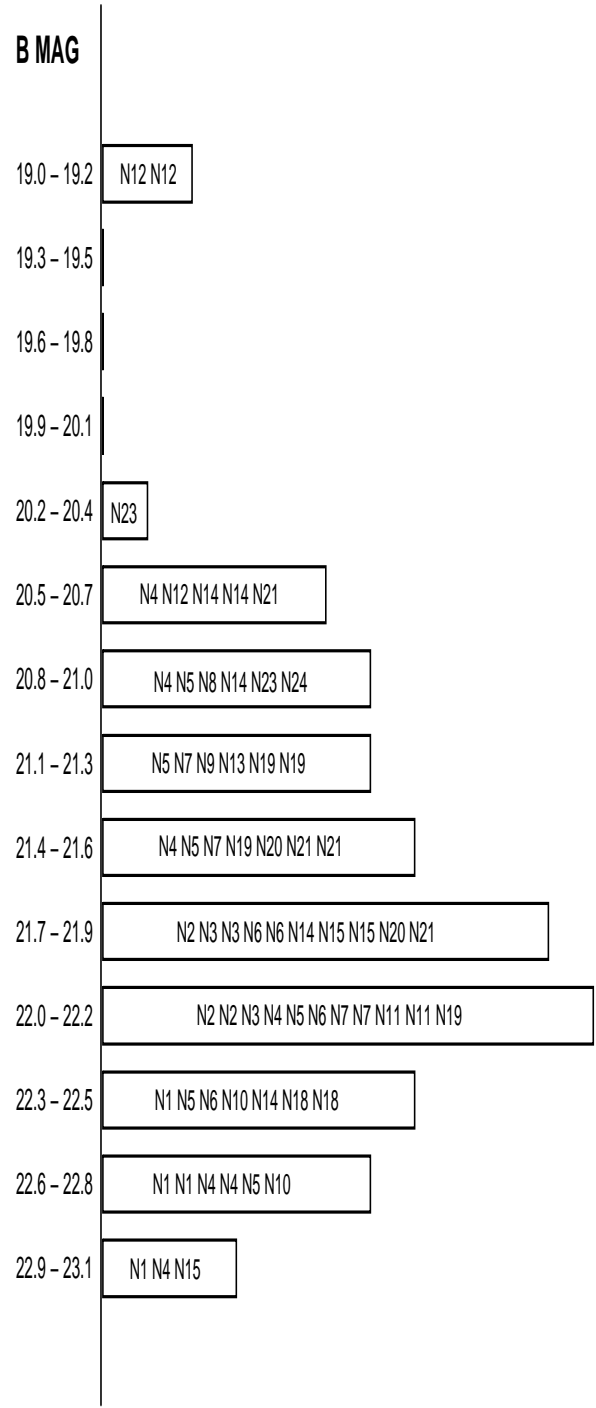
\end{document}